\documentclass[11pt,a4paper]{article}

\usepackage{amsfonts}
\usepackage[dvips]{graphicx}
\usepackage{amsthm}
\usepackage{epsfig}
\usepackage{t1enc}
\usepackage{amssymb}
\usepackage{latexsym}
\usepackage{bm}

\title{\bf Euclidean action
\\
for vacuum decay in a de Sitter universe}
\author{V. Balek\footnote{e-mail address: balek@fmph.uniba.sk}\ \ and
M.Demetrian\footnote{e-mail address: demetrian@sophia.dtp.fmph.uniba.sk}
\\
{\it Department of Theoretical Physics,
Comenius University, Mlynsk\'a dolina}
\\
{\it 842 48 Bratislava, Slovakia}}

\begin{document}
\maketitle
\maketitle\abstract

{The behavior of the action of the instantons describing vacuum decay
in a de Sitter is investigated. For a near-to-limit instanton (a
Coleman-de Luccia instanton close to some Hawking-Moss instanton)
we find approximate formulas for
the Euclidean action by expanding the scalar field and the metric of
the instanton in the powers of the scalar field amplitude. The order
of the magnitude of the correction to the Hawking-Moss action
depends on the order of the instanton (the number of crossings of
the barrier by the scalar field): for instantons of odd and even
orders the correction is of the fourth and third order in the scalar
field amplitude, respectively. If a near-to-limit instanton of the
first order exists in a potential with the curvature at the
top of the barrier greater than 4 $\times$ (Hubble constant)$^2$,
which is the case if the fourth derivative of the potential at the
top of the barrier is greater than some negative limit value,
the action of the instanton is less than the Hawking-Moss action and,
consequently, the instanton determines the outcome of the vacuum
decay if no other Coleman-de Luccia instanton is admitted by
the potential.
A numerical study shows that for the quartic potential the physical
mode of the vacuum decay is given by the Coleman-de Luccia instanton
of the first order also in the region of parameters in which the
potential admits two instantons of the second order.}


\section{Introduction}

The vacuum decay in a de Sitter universe has been considered as a
mechanism of transition to a Friedman universe in the scenario of
old inflation \cite{old}, and as a triggering mechanism in the
scenario of open inflation \cite{bgt}. In both scenarios,
the instability of the false vacuum results in the
formation of rapidly expanding bubbles containing the scalar
field on the true-vacuum side of the potential barrier. The process
is described by the Coleman-de Luccia (CdL) instanton \cite{cdl}.
Another kind of vacuum decay is mediated by the Hawking-Moss (HM)
instanton \cite{hmo} and proceeds in such a way that the field
jumps on the top of the barrier inside a horizon-size domain. (For
this interpretation, see \cite{lst}.) Recently this process was
proposed as a starting point for the inflation driven by the trace
anomaly \cite{hhe}.

An important quantity characterizing the instanton is
its action. It determines the probability of an instanton-induced
transition per unit spacetime volume: if we denote the action by
$B$, the probability is proportional to exp$(-B)$. Thus, if more
instantons exist for a given potential, the one with the least
action prevails. In addition, the value of the action may be
relevant to the question whether the instanton describes a well
defined quantum mechanical transition at all. The transition
takes place only if the instanton admits no negative modes of the
perturbation superimposed on it; and according to \cite{tan}, such
modes are certainly not present only for the instanton with the
least action. If negative modes exist, the preexponential
factor in the expression for the probability diverges, which
indicates that the wave function is not concentrated in a narrow
tube around the instanton solution as it should be. (For an
analogy in ordinary quantum mechanics, see \cite{2Dt}.) Thus, the
instanton with the least action, in addition to being the
most probable, may be the only one with a physical meaning.

The values of the action of CdL instantons were discussed in the
thin wall approximation in \cite{cdl} and \cite{par}, and computed
for a one-parametric class of quartic potentials admitting an instanton
of the first order (an instanton with the scalar field crossing
the top of the barrier just once) in \cite{sam}. Furthermore, the
behavior of the action of CdL instantons was analysed in the
approximation of negligible back reaction in \cite{sst}; the properties
of a near-to-limit instanton (a CdL instanton close to some HM
instanton) were studied with the back reaction included in \cite{sta};
and the instanton with the least action was identified among the
instantons of odd orders in \cite{tan}. Here we link up to
this research and develop it further. In sections \ref{sec:B2} and
\ref{sec:B34} we find a formula for the action of a near-to-limit
instanton with the back reaction included, and in section \ref{sec:Ex}
we compute the instanton action for a class of quartic potentials with
a subclass admitting instantons of the second order.
Thus, in the first two sections we generalize the result of
\cite{sst} and complete the results of \cite{sta}, while in the
last section we extend the results of \cite{tan} and \cite{sam}.



\section{The second order contribution to the action}
\label{sec:B2}

Consider a compact O(4) symmetric 4D space and an O(4) symmetric
scalar field $\phi$ living in it. The system is completely described
by the functions $\phi (\rho)$ and $a (\rho)$, where $\rho$ is
the distance from the "north pole" (the radius of 3-spheres
of homogeneity measured from the center) and $a$ is the scale
parameter (the radius of 3-spheres of homogeneity determined from
the circumference). The compactness implies that $\rho$ runs from
zero to some $\rho_f$ and $a = 0$ at both $\rho = 0$ and $\rho_f$.
The Euclidean action is ($\hbar = c = 1$)
\begin{equation}
B = 2\pi^2 \int_0^{\rho_f} \left[ \left(\frac 12 \dot \phi^2 + V \right) a^2
- \frac 1\kappa (\dot a^2 + 1) \right] a\ d\rho,
\label{eq:B}
\end{equation}
where the overdot denotes differentiation with respect to $\rho$,
$V$ is the effective potential of the scalar field and $\kappa = 8\pi
G/3$. The extremization of the action yields
\begin{equation}
\ddot \phi + 3 \frac {\dot a}a \dot \phi = \partial_\phi V, \ \
\ddot a = - \kappa (\dot \phi^2 + V) a.
\label{eq:diff}
\end{equation}
The equation for $a$ is obtained by differentiating the
equation
$$\dot a^2 = \kappa \left( \frac 12 \dot \phi^2 - V \right) a^2 +
1,$$
and inserting for $\ddot \phi$ from the equation for $\phi$.
(In fact, when putting the variation of $B$ equal to zero we obtain
the first order equation for $a$ with an additional term
$Ca^{-2}$, where $C$ is an integration constant. To suppress
this term one has to introduce the lapse function into the
expression for $B$ and perform variation with respect to it.)
Consider functions $\phi$ and $a$
extremizing the action and such that the action is finite for them.
Functions with these properties may be found, in principle,
by solving equations (\ref{eq:diff}) with the boundary conditions
\begin{equation}
\dot \phi (0) = \dot \phi (\rho_f) = 0,\ \ a(0) = 0,\ \
\dot a (0) = 1.
\label{eq:bound}
\end{equation}
Since $\phi$ and $\dot \phi$ are finite at the points $\rho = 0$
and $\rho_f$, $\dot a$ and $\ddot a$ are finite at these points,
too. Consequently, we may simplify the integral in (\ref{eq:B}) by
replacing
$$a\dot a^2 \to - \frac 13 a\dot a^2 - \frac 23 a^2\ddot a$$
and using both equations for $a$ to obtain
\begin{equation}
B = - \frac {4\pi^2}{3\kappa} \int_0^{\rho_f} a\ d\rho.
\label{eq:Bsimp}
\end{equation}
This expression is appropriate for numerical calculations but
when expanding the action into the powers of the scalar field
amplitude it is advisable to use expression (\ref{eq:B}) instead.
Then one does not need to compute $a$ up to the same order
of magnitude as $B$; as we shall see, one gets along with the zeroth
and second order contributions to $a$ when computing the third and
fourth order contributions to $B$, respectively.


Consider a potential with the global minimum, or the true vacuum, at
$\phi = 0$, a maximum at $\phi = \phi_M > 0$ and a local minimum, or
the false vacuum, at $\phi = \phi_m > \phi_M$. Suppose, furthermore,
that the potential equals zero at the true vacuum. There are two
trivial finite-action solutions to equations (\ref{eq:diff}) with
a compact 4-geometry,
$$\phi = \phi_M, \ \ \ a = H_M^{-1} \sin (H_M\rho), \ \ \
0 < \rho < \pi H_M^{-1},$$
and
$$\phi = \phi_m, \ \ \ a = H_m^{-1} \sin (H_m\rho), \ \ \
0 < \rho < \pi H_m^{-1},$$ where $H_M$ and $H_m$ are the values of
$H = \sqrt{\kappa V}$ (the Hubble constant corresponding to a
given value of $\phi$) at $\phi = \phi_M$ and $\phi_m$
respectively. Both solutions consist of a 4-sphere and a constant
scalar field. The first solution is the HM instanton; the second
solution describes a "zero quantum transition" hence it may be
regarded as a reference solution. In addition, the potential may
admit nontrivial finite-action solutions to equations
(\ref{eq:diff}) called the CdL instantons. These solutions may be
classified according to how many times the scalar field crosses
the barrier: if the number of crossings is $l$, the instanton may
be called "the CdL instanton of the $l$th order". Denote $\xi =
\partial^2_\phi V_M/H_M^2$. For $- \xi$ assuming one of the
critical values
\begin{equation}
\lambda_l = l(l + 3) = 4, 10, \ldots \ \ \mbox{for}\ \ l = 1, 2,
\ldots,
\label{eq:lambdal}
\end{equation}
the potential admits an approximate CdL instanton consisting of a
4-sphere with the radius $H_M^{-1}$ and a scalar field
proportional to the $l$th harmonic on the 4-sphere
\cite{jst}. This solution may be called "the limit instanton
of the $l$th order". If $\xi$ is close to $- \lambda_l$, at least
from one side, the potential should admit a CdL instanton
close to the limit instanton with some small scalar field amplitude.
In a one-parametric family of potentials, this near-to-limit
instanton should approach the HM instanton as $\xi$ approaches
$- \lambda_l$.



Introduce dimensionless variables
$$\chi = H_M \rho,\ \ u = \frac 1A \Delta \phi,\ \ v = H_M a,$$
where $\Delta \phi = \phi - \phi_M$ and $A = \Delta \phi (0)$,
and write the potential as
$$V = H_M^2 \left[\frac 1\kappa + A^2 \left(\frac 12 \xi u^2 +
\omega\right)
\right].$$
The expression for the action, when rewritten into the dimensionless variables,
is
\begin{equation}
B =  \frac {2\pi^2}{H_M^2} \int_0^{\chi_f} \left[A^2
\left(\frac 12 {u'}^2 + \frac 12 \xi u^2 + \omega\right) v^2 - \frac
1\kappa ({v'}^2 - v^2 + 1) \right] v\ d\chi,
\label{eq:Be}
\end{equation}
and the instanton equations are
\begin{equation}
u'' + 3 \frac {v'}v u' - \xi u = \partial_u \omega, \ \
v'' + v = - A^2 \left({u'}^2 + \frac 12 \xi u^2 + \omega\right) v,
\label{eq:diffe}
\end{equation}
where $\chi_f = H_M \rho_f$ and the prime denotes differentiation
with respect to $\chi$. Denote $s = \sin \chi$ and $c = \cos \chi$.
For the HM instanton $u = 0$, $v = s$ and $\chi_f = \pi$,
which yields the action
\begin{equation}
B_M = - \frac {8\pi^2}{3\kappa H_M^2}.
\label{eq:BM}
\end{equation}
For the limit instanton, $u$ obeys the first equation
(\ref{eq:diffe}) with $v = s$, $\xi = - \lambda_l$ and suppressed
right hand side,
$${\cal D} u = 0,\ \ \ {\cal D} = \frac {d^2}{d^2\chi} -
3\mbox{cotan} \chi\ \frac d{d\chi} + \lambda_l.$$
The solutions are O(4) symmetric spherical harmonics in 5D or,
up to normalization, the Gegenbauer polynomials with the parameter
3/2,
\begin{equation}
P_l = c,\ \frac 14 (5c^2 - 1),\ \ldots \ \ \mbox{for}\ \ l = 1, 2,
\ldots
\label{eq:Pl}
\end{equation}
When constructing a near-to-limit instanton one introduces
expansions $\xi = \xi_0 + A \xi_1 + A^2 \xi_2 + \ldots$,
$u = A u_1 + A^2 u_2 + A^3 u_3 + \ldots$ and
$v = v_0 + A^2 v_2 + A^3 v_3 + \ldots$
with $\xi_0 = - \lambda_l$, $u_1 = P_l$ and $v_0 = s$, and uses the
expansion
$$\omega = \frac 16 \eta A u^3 + \frac 1{24} \zeta A^2 u^4 +
\ldots,$$
where $\eta = \partial^3_\phi V_M/H_M^2$ and $\zeta =
\partial^4_\phi V_M/H_M^2$.
>From the expanded equations for $u$ and $v$ one obtains the functions
$u_1$, $u_2$, $\ldots$ and $v_0$, $v_2$, $\ldots$, and after
inserting them into the expression for $B$ one finds the
coefficients $B_0$, $B_2$, $\ldots$ in the expansion
$B = B_0 + A^2 B_2 + \ldots$
In particular, from $v_0 = s$ one obtains $B_0 = B_M$.


Let us prove that the term proportional to $B_2$ in the expansion
of $B$ vanishes so that $B = B_M$
up to the order $A^2$. For $B_2$ we have
\begin{equation}
B_2 =  \frac {2\pi^2}{H_M^2} \int_0^\pi \left\{\frac 12
({P_l'}^2 - \lambda_l P_l^2) s^3 - \frac 1\kappa [2sc v_2' +
(c^2 - 3s^2 + 1) v_2] \right\} d\chi.
\label{eq:B20}
\end{equation}
(We have not included here a term arising from the
shift of the upper limit of integration since such terms do not
appear before the order $A^4$.)
The first part, with the terms quadratic in $P_l$, may be rewritten
using integration by parts to an integral of $P_l {\cal D}
P_l$, hence it vanishes. The second part, with the terms linear in
$v_2$, may be rewritten using integration by parts to an integral
of ${\cal A} v_2$ with the coefficient of proportionality
$${\cal A} = - (2sc)' + c^2 - 3s^2 + 1 = 0,$$
hence it vanishes, too.
This can be seen also without an explicit calculation. The second
part of $B_2$ is just what we obtain if we perform the variation
of $B_0$ with respect to $v_0$
and put $\delta v_0 = v_2$, thus ${\cal A} = 0$ is in fact the
equation determining $v_0$. As a result we obtain
\begin{equation}
B_2 = 0.
\label{eq:B2}
\end{equation}


\section{Higher order contributions to the action}
\label{sec:B34}

The first higher order term in the expansion of $B$ which may be nonzero
is proportional to $B_3$. The terms in $B_3$ linear in $u_2$ and $v_3$ can
be rewritten to be proportional to ${\cal D} P_l$ and $\cal A$,
thus they vanish and we are left with the expression
\begin{equation}
B_3 =  \frac {\pi^2}{H_M^2} \left(\xi_1 I_2 + \frac 13 \eta I_3 \right),
\ \ \ I_k = \int_0^\pi P_l^k s^3 d\chi.
\label{eq:B30}
\end{equation}
To determine $\xi_1$, consider the equation for $u_2$
\begin{equation}
{\cal D} u_2 = \xi_1 P_l + \frac 12 \eta P_l^2.
\label{eq:ph2}
\end{equation}
The task simplifies considerably by two observations: first, the left
hand side, when expanded into the system of harmonics $P_0 = 1$,
$P_1$, $P_2$, $\ldots$, does not contain a term proportional to
$P_l$ (this is easily seen if one expands $u_2$ into the harmonics
and makes use of the fact that the operator ${\cal D}$, when applied
on $P_{l'}$, reproduces it with a factor that is zero for $l' = l$);
and second, the harmonics are orthogonal with the weight $s^3$. Thus,
if we multiply equation (\ref{eq:ph2}) by $P_l$ and integrate it with
the weight $s^3$ we obtain zero on the left hand side. As a result
we have
$$\xi_1 I_2 + \frac 12 \eta I_3 = 0,$$
hence
$$B_3 = - \frac {\pi^2}{6H_M^2} \eta I_3.$$
The $l$th harmonic is a linear combination of $c^l$, $c^{l - 2}$,
$\ldots$, thus $P_l^3 s^3$ is an odd function on the interval $(0, \pi)$
and $I_3$ vanishes if $l$ is odd. On the other hand,
from the general formula for the integral of the product of three
Gegenbauer polynomials \cite{vil} it follows that $I_3$ is
nonzero if $l$ is even. Thus, for even values of $l$ both $\xi_1$
and $B_3$ are nonzero and the quantities
$\Delta \xi = \xi + \lambda_l$ and $\Delta B = B - B_M$
are of the order $A$ and $A^3$ respectively. Explicitly,
\begin{equation}
\Delta \xi \doteq - \frac {I_3}{2I_2} \eta A,
\label{eq:Dxi1}
\end{equation}
and
\begin{equation}
\Delta B \doteq - \frac{\pi^2 I_3}6 \ \frac {\eta A^3}{H_M^2}.
\label{eq:B3}
\end{equation}
The actual small parameter of the theory is not $A$ but
$\Delta \xi$, therefore we have to interpret the first equation
as an approximate expression for $A$ in terms of $\Delta \xi$.
For $l = 2$, $I_2 = 2/21$ and $I_3 = 2/63$,
thus the dimensionless coefficients in the expressions for $\Delta
\xi$ and $\Delta B$ are 1/6 and $\pi^2/189$ respectively. Note that
we may avoid the computation of $I_3$ if we express $B_3$ in terms
of $\xi_1$ and determine $\xi_1$ from the requirement that the term
proportional to $P_l$ on the right hand side of (\ref{eq:ph2})
vanishes.

To summarize, the correction to the HM action of order $A^3$
vanishes for instantons of odd orders, but
it is nonzero for instantons of even orders provided $\eta$
is nonzero. If we change the parameters of the potential
so that $\xi$ changes while $\eta$ stays fixed, the character of the
instanton changes as $\xi$ crosses the value $- \lambda_l$.
If, say, $\eta > 0$, for $\xi < - \lambda_l$ the function $\phi$
starts and ends to the right of $\phi_M$ (the instanton is "right
handed"), while for $\xi > - \lambda_l$ the function $\phi$
starts and ends to the left of $\phi_M$ (the
instanton is "left handed"). No matter what the handedness of the
instanton, the action of the instanton is less than the HM action
if $\xi < - \lambda_l$ and greater than the HM action if $\xi
> - \lambda_l$.


To complete the analysis we have to compute $B_4$ for odd values
of $l$.
First, just as when we were computing $B_3$, we get rid of a
large portion of the integrand by rewriting it in terms of ${\cal D}
P_l$ and $\cal A$; in this way we remove the terms
proportional to $u_3$ and $v_4$. Before we present the
remaining terms let us mention a new point which arises in this
order of expansion of $B$. One may expect that the value of
$v_2$ at $\chi = \pi$ is nonzero and indeed, for $l = 1$ one
finds $\delta \equiv v_2 (\pi) = - \pi \kappa/8$ (see the
appendix). Consequently, $\chi_f$ differs from $\pi$
by a quantity of order $A^2$. As we shall see, this
leads to additional terms in $B_4$ so that not the whole $B_4$ is
stemming from the expansion of the integrand of $B$. After
suppressing the terms proportional to $u_3$ and $v_4$ and
integrating by parts we arrive at
\begin{equation}
B_4 = \frac {\pi^2}{H_M^2} \left\{ \xi_2 I_2 + \frac 1{12} \zeta I_4
+ \int_0^\pi \left[ u_2 (- {\cal D} u_2 + P)
s^3 + v_2 \left(- \frac 2\kappa {\cal B} v_2 + Q \right)
\right] d\chi + b_4 \right\},
\label{eq:B40}
\end{equation}
where
$$P = \eta P_l^2, \ \ Q = 3 ({P_l'}^2 - \lambda_l P_l^2)
s^2,\ \ {\cal B} = - s \frac {d^2}{d^2\chi} - c \frac d{d\chi} -
2s,$$
and $b_4$ is a boundary term consisting of two parts, one
coming from the shift in $\chi_f$ and the other coming from the
integration by parts which has to be performed when deriving the
term with the operator $\cal B$. The expression for $B_4$ simplifies
further if we exploit the equations for $u_2$ and $v_2$,
\begin{equation}
{\cal D} u_2 = \frac 12 P,\ \ \ {\cal B} v_2 =
\frac 14 \kappa Q.
\label{eq:ph2al2}
\end{equation}
Both equations are most easily derived by performing the variation
of $B_4$ and using the fact that the operators $s^3 \cal D$
and $\cal B$ are symmetric;
however, they may be obtained by expanding the exact equations for
$u$ and $v$ as well. The former equation
is identical to (\ref{eq:ph2}) provided $\xi_1 = 0$, which is what we
presently assume, and the latter equation is a linear combination of
the two equations for $v_2$ presented in the appendix. If we make use
of equations (\ref{eq:ph2al2}) and of the definition of $P$, we find
that the integral in $B_4$ reduces to
$$\frac 12 \eta J + \frac 1\kappa \int_0^\pi Q v_2\ d\chi,
\ \ \ J = \int_0^\pi P_l^2 u_2 s^3\ d\chi.$$

Let us now show that the boundary term in $B_4$ is zero. We can find
the correction to the upper limit of integration by noticing that
the derivative of $v_0$ at $\chi = \pi$ is $- 1$, so that to
compensate for $\Delta v \doteq A^2 v_2$ in the neighborhood of
$\chi = \pi$ we have to shift $\chi_f$ by $\Delta = A^2
\delta$. The contribution to $b_4$ arising from this shift is
$$b_4^{shift} = A^{-4} \times\ \mbox{the leading term in}\
\int_\pi^{\pi + \Delta} 2 (F_0 + A^2 F_2)\ d\chi,$$
where $F_0$ and $F_2$ are the integrands in $B_0$ and $B_2$,
$$F_0 = \ldots - \frac 1\kappa (c^2 + 1) s,\ \ \
F_2 = \ldots - \frac 1\kappa (c^2 + 1)v_2.$$
(The omitted terms are irrelevant for the present discussion.)
The leading term in the integral of $F_0$ equals $F_0' (\pi)
\Delta^2/2 = \Delta^2 /\kappa$, since $F_0 (\pi) = 0$, and
the leading term in the integral of $F_2$ equals $F_2 (\pi) \Delta =
- 2\delta \Delta/\kappa$, hence
$$b_4^{shift} = - \frac {2\delta^2}{\kappa}.$$
On the other hand, from the expression of the relevant part of $B$ given
in the appendix it follows that the contribution to
$b_4$ coming from the integration by parts is
$$b_4^{int} = \frac {2\delta^2}{\kappa}.$$
Putting this together we find
\begin{equation}
b_4 = 0.
\label{eq:b4}
\end{equation}

Next we have to write down the equation for $u_3$ in
order to determine $\xi_2$. When constructing this equation we
encounter a subtlety that is again connected with the presence
of $\delta$. If $\delta$ is nonzero, $v_2$
is not small with respect to $s$ in the vicinity of $\chi = \pi$
and the factor $\dot a/a = H_M v'/v$ in the equation
for $\phi$ cannot be treated perturbatively. In fact, if we formally
expand $v'/v$ up to the order $A^2$,
$$\frac {v'}v \doteq \mbox{cotan}\chi + A^2 q,\ \ \
q =  \left( \frac{v_2}s \right)',$$
we can see that the correction term even diverges relatively to the
zeroth order term in $\chi = \pi$. (It behaves like $- \Delta/
\epsilon^2$ for $\epsilon = \pi - \chi \to 0$, while the zeroth
order term behaves like $1/\epsilon$.)
To fix that, note that the term $\hat \delta$ in $v_2$ which
is nonzero at $\chi = \pi$ and therefore is responsible for the
nonzero value of $\delta$ equals $- \delta \chi v_0'/\pi$
(see the appendix), thus $\hat \delta$ may be
absorbed into $v_0$ by redefining $\chi \to (1 - \Delta/\pi)\chi$.
Such procedure is well known in the perturbation theory of
an anharmonic oscillator where the terms of the type $\hat \delta$
are called "resonance terms" \cite{lli}. The corrected equation for
$v_3$ is obtained in such a way that we replace $q$ by
$q^{red}$ ("red" standing for "reduced"), defined as $q$ with $v_2$
replaced by $v_2^{red} = v_2 - \hat \delta$, and add to $\xi_2$
the term $\kappa$ arising from the redefinition of $\chi$
in the equation for $u_1$. However,
these corrections, necessary as they are when one computes $v_3$, are
superfluous if one is interested only in the integral of the
equation for $v_3$ with the weight $s^3$. The point is that
the weight washes out the singularity in $q$, therefore the reverse
transformation $\chi \to (1 + \delta/\pi)\chi$ by which one passes
from the corrected equation to the original one may be treated on
equal footing with the regular transformations (that is,
transformations that do not move the point $\chi = \pi$). To confirm
this, one may check that the two corrections mentioned above cancel
when the equation is integrated over $\chi$ with the weight $s^3$.
The uncorrected equation for $v_3$ reads
\begin{equation}
{\cal D} v_3 = \xi_2 P_l + \eta _1 P_l u_2 + \frac 16 \zeta
P_l^3 - 3q P_l',
\label{eq:ph3}
\end{equation}
and by the same argument as we have used when calculating $B_3$ we
can deduce from it the identity
$$\xi_2 I_2 + \eta J + \frac 16 \zeta I_4 + \int_0^\pi 3 q P_l P_l'
s^3\ d\chi = 0.$$
Finally, by integrating by parts and using the equation for $P_l$
it may be shown that the integrand in the last term can be
replaced by $- Q v_2$. This completes the proof that in our initial
expression for $B_4$ the sum of all terms following $\xi_2 I_2$
equals $- \xi_2 I_2/2$, and
$$B_4 = \frac {\pi^2} {2H_M^2} I_2 \xi_2.$$
Similarly as when we were computing $B_3$ we can now obtain
approximate expressions for $\Delta \xi$ and $\Delta B$
in terms of $A$. Again, we may
determine $\xi_2$ also by analysing the equation for $v_3$; however,
for that purpose the corrected equation has to be used, with $q$
replaced by $q^{red}$ and $\xi_2$ by $\xi_2 + \kappa$. Let us present
the results for the most important case $l = 1$. In this case,
\begin{equation}
\Delta \xi \doteq - \frac 1{14} \left(\frac 1{12}\eta^2 + \zeta +
32 \kappa\right) A^2
\label{eq:Dxi2}
\end{equation}
and
\begin{equation}
\Delta B \doteq \frac{2\pi^2}{15}\ \frac{\Delta \xi A^2}{H_M^2}.
\label{eq:B4}
\end{equation}
The expression for $\Delta \xi$ implies that a near-to-limit instanton
exists only if $- \xi$ approaches the value $\lambda_1 = 4$ from a
given side. Which side, that depends on the parameters $\eta$ and $\zeta$.
Define
\begin{equation}
\zeta_{crit} = - \frac 1{12} \eta^2 - 32 \kappa.
\label{eq:zcrit}
\end{equation}
If $\zeta > \zeta_{crit}$ the instanton exists for $\xi < - 4$,
while if $\zeta < \zeta_{crit}$ the instanton exists for $\xi > -
4$. The action of the instanton is less than the HM action
in the former case and greater than the HM action in the latter case.

The expression for $\Delta \xi$ is identical to that obtained in
\cite {sta}, and if we suppress $\kappa$ in the expression for $\Delta \xi$
and insert for $H_M$, $\eta$ and $\zeta$ the values for the quartic potential,
the resulting expression for $\Delta B$ is identical to that
derived in \cite{sst}. Our formula for $\Delta B$ is consistent
also with the behavior of the first perturbation mode of the
near-to-limit instanton established in \cite{sta}, since this mode
is positive (and, consequently, contributes to the
action by a positive quantity) in the same range of
parameters in which we found that $B_M - B$ is positive.



\section{The instanton with the least action}
\label{sec:Ex}
\begin{figure}[ht]
\centerline{\includegraphics[width=14cm]{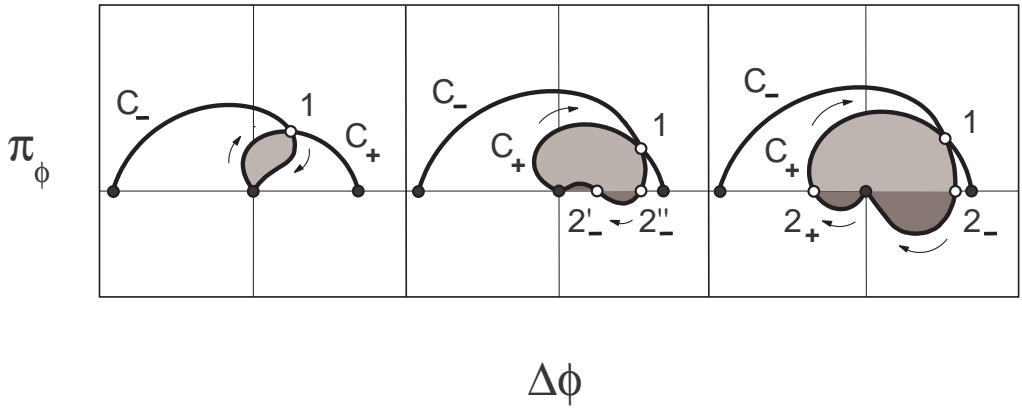}}
\caption{Phase diagrams for the solutions of the Euclidean theory}
\label{fig:cpm}
\end{figure}

The question which instanton has the least action is addressed
in \cite{tan} where it is shown that among
the instantons of odd orders, the one with the least action is
necessarily of the first order. The proof is based on the
analysis of a graph that may be regarded as a phase
diagram for the solutions of the Euclidean theory. One introduces {\it
noninstanton} solutions $\phi_- (\rho)$ and $\phi_+ (\rho)$ starting
to the left and to the right of $\phi_M$, and assigns to them
the values of $\Delta \phi$ and $\pi_\phi = 2\pi^2 a^3 \dot \phi$
at such $\rho$ at which the function $a (\rho)$ reaches
maximum. In this way one obtains two oriented curves
in the $(\Delta \phi, \pi_\phi)$ plane: the curve $C_-$ starting
at the point $(-\phi_M, 0)$ (the true vacuum) and ending at the origin
(the top of the barrier), and the curve $C_+$ starting at the origin
and ending at the point $(\phi_m - \phi_M, 0)$ (the false vacuum).
An instanton of an odd order is represented by an intersection
of the curves $C_+$ and $C_-$, and its "net action" $\Delta B$
equals the area inside a loop passing from the origin to the
intersection and back, first along the curve $C_+$ and then along
the curve $C_-$. (The area is regarded here as the
$z$-component of the vector $\oint {\bf r} \times d{\bf r}$, where $\bf
r$ is the radius vector in the $(x, y)$ plane. According to this
definition, the area inside a given curve is positive if the curve is
oriented counterclockwise and negative if the curve is oriented
clockwise.) As to the instantons of even orders, they are represented
by the points where the curves $C_-$ (for left handed instantons)
and $C_+$ (for right handed instantons) intersect the axis $\Delta
\phi$, and their net action
equals twice the area inside a loop passing from the origin to the
intersection and back, first along the curve $C_-$ and then along the
axis $\Delta \phi$ if the instanton is left handed, and first along the
axis $\Delta \phi$ and then along the curve $C_+$ if the instanton is
right handed. Three typical phase diagrams are depicted in fig.
\ref{fig:cpm}.
All diagrams contain an instanton of the first order denoted
by $1$; besides, the middle diagram contains two left handed
instantons of the second order denoted by $2_-'$ and $2_-''$, and
the right diagram contains one left handed and one right handed
instanton of the second order denoted by $2_-$ and $2_+$. The net
action of the instanton of the first order equals the area of the
light grey patch
in the left diagram; the area of the the light grey patch and the lower
dark grey patch in the middle diagram; and the area of the light
grey patch and both dark grey patches in the right diagram. The limit
curves are in all three cases oriented clockwise, therefore the area is
negative and the action of the instanton is less than the HM action.
The net action of the instantons of the second order equals twice the
area of the corresponding dark grey patches; consequently, the net
action is negative for both $2_-$ and $2_+$, positive for $2_-'$, and
either negative or positive, depending on which of the two dark grey
patches prevails, for $2_-''$. There seems to be no general argument
restricting the area of the dark grey patches in comparison
with the light grey one, therefore the less of the actions of the two
instantons of the second order may be also less than the action of the
instanton of the first order. As seen from the geometry of the patches,
one can only state
that the action of the instanton of the first order is less than the
{\it mean} action of the instantons of the second order.

The phase diagrams suggest that if we take into consideration
the instantons of even orders (including the HM instanton which may be
regarded as the instanton of the zeroth order), the instanton with the
least action may not stay of the first order. We have seen
that if two instantons of the second order exist in addition to one
instanton of the first order, the instanton with the least action may
one of them. In a similar way one may prove that if there are
two instantons of the first order and none of a higher order (which
occurs if the curves $C_-$ and $C_+$ are intertwined next to the origin
and do not intersect the axis $\Delta \phi$ somewhere), and if both
instantons have approximately the same action
(which occurs if the two intersections of the curves $C_-$ and $C_+$
outside the origin are close to each other), the instanton with the
least action is necessarily the HM instanton. For a typical potential,
however,
one expects the instanton of the first order to prevail even in the
presence of instantons of even orders. To support this opinion, we
present here a numerical study for a certain class of quartic
potentials.

The quartic potential is usually regarded as depending
on three parameters, however the parameter $m$ whose square appears
in front of $\phi^2$ can be removed from the
theory if one makes use of rescalings $a \to a/m$ and $\rho \to
\rho/m$ in the equations for $\phi$ and $a$. The rescaled potential is
\begin{equation}
V = \frac 12 \phi^2 - \frac 13 \delta \phi^3 + \frac 14 \lambda
\phi^4.
\label{eq:Vqu}
\end{equation}
The potential has the
desired form only for the parameter $\delta$ running from $\delta_m =
2 \sqrt{\lambda}$ to $\delta_M = 3 \sqrt{\lambda/2} \doteq 1.06
\delta_m$. For numerical calculations we have chosen the value of $\delta$
in the middle of this interval,
$$\Delta \equiv \frac {\delta - \delta_m}{\delta_M - \delta_m} =
0.5.$$
The parameter $\xi$ may be expressed in terms of $\delta$ and
$\lambda$ as ($G = 1$)
\begin{equation}
\xi = - \frac {9\lambda}\pi\ \frac {X (1 + X)} {(1 - X)(1 + 3X)},
\ \ \ X = \sqrt{1 - \frac{4\lambda}{\delta^2}}.
\label{eq:xiqu}
\end{equation}
Note that $X$ depends on $\delta$ and $\lambda$ only through $\Delta$,
so that if one keeps $\Delta$ fixed $\xi$ becomes proportional to
$\lambda$.

\begin{figure}[ht]
\centerline{\includegraphics[width=11.5cm]{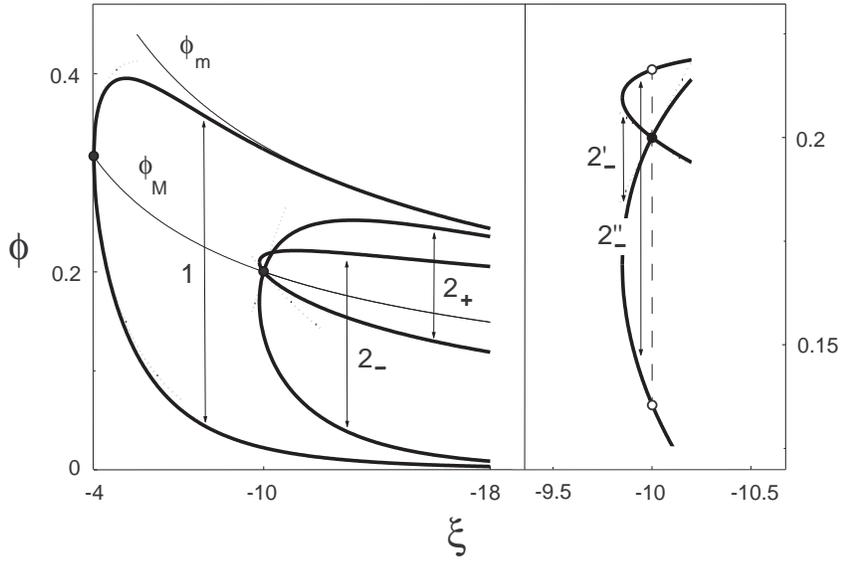}}
\caption{Range of values of the scalar field
for instantons in the quartic potential}
\label{fig:phi}
\end{figure}

\begin{figure}[ht]
\centerline{\includegraphics[width=11.5cm]{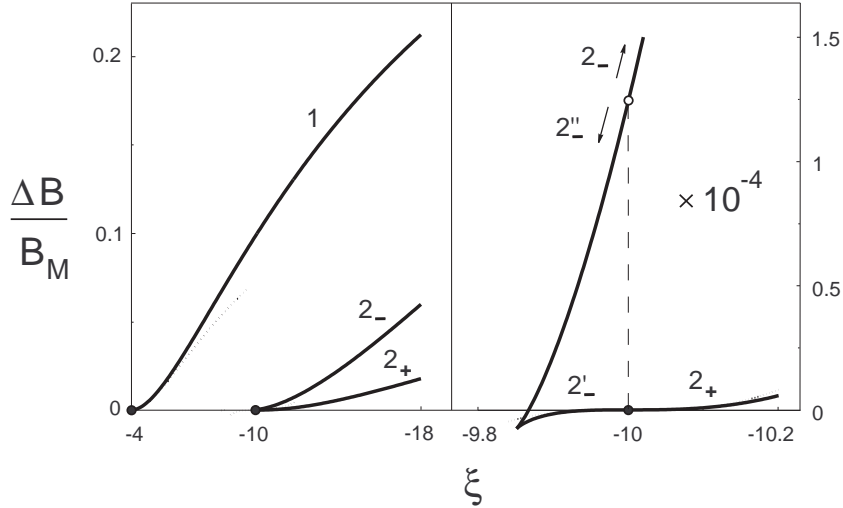}}
\caption{Action of instantons in the quartic potential}
\label{fig:DB}
\end{figure}

The results are summarized in fig. \ref{fig:phi} and \ref{fig:DB}.
The characteristics of instantons obtained numerically by
solving the equations (\ref{eq:diff}) are depicted by solid lines,
and those computed from the formulas (\ref{eq:Dxi2}) and (\ref{eq:B4})
in the neighborhood of $\xi = - 4$, and from the formulas
(\ref{eq:Dxi1}) and (\ref{eq:B3}) in the neighborhood of $\xi = - 10$,
are depicted by dotted lines. As $\xi$ decreases, the phase diagram
passes through the three regimes visualised in
fig. \ref{fig:cpm}. They occur, from the left to the right,
for $\xi$ running from $- 4$ to some $\xi_{crit}$ greater
than $- 10$; for $\xi$ running from $\xi_{crit}$ to $- 10$; and
for $\xi$ running from $- 10$ to $- 18$. The behavior of
instantons near the limit values of $\xi$ is determined by the fact
that for the quartic potential both $\eta$ and $\zeta$ are positive.
As a result, the first order instantons exist only for $\xi < - 4$
and the near-to-limit second order instantons are right handed for
$\xi < - 10$ and left handed for $\xi > - 10$.

In fig. \ref{fig:phi} the limit values of $\phi$ are plotted as
functions
of the parameter $\xi$. The graph to the right provides a
magnified view of a narrow segment of the graph to the left.
For comparison, the location of the top of the barrier and of
the false vacuum is shown in the graph to the left.

In fig. \ref{fig:DB} the ratio $\Delta B/B_M$ is plotted as a
function of the parameter $\xi$.
Again, we can see a tiny segment of the graph
to the left displayed in more detail in the graph to the right.
The values of $\Delta B/B_M$ in the latter graph are to
be multiplied by $10^{-4}$. The HM action is negative, hence
in the region above the horizontal axis there holds $\Delta B
< 0$.


\section{Conclusion}
\label{sec:Con}

The behavior of the action of CdL instantons has been investigated
in two complementary ways, analytically for a general potential
with the value of $\xi$ (the rescaled second derivative of $V$ at
$\phi = \phi_M$) near to one of the limit values $- 4, - 10, - 18,
\ldots$, and numerically for a class of quartic potentials with the
value of $\xi$ running from $- 4$ to $- 18$. The analytical study
implies that
the properties of instantons of odd and even orders are in many
respects different. While the instantons of even orders exist for
$\xi$ on both sides of the limit value, the instantons of odd orders
exist only for $\xi$ either to the left or to the right of the
limit value, depending on the values of $\eta$ and $\zeta$
(the rescaled third and fourth derivative of $V$ at $\phi = \phi_M$).
Also the dependence of the action on the scalar field amplitude
is different for instantons of even and odd orders. However, the
two kinds of instantons have one property in common: their
action is less than the HM action if the value of $\xi$
is less than the limit value, that is if the potential
is more curved at the top of the barrier than the potential for
which the near-to-limit instanton shrinks to the top of the barrier,
and vice versa. (Note that this may be proven also without an
explicit computation, by investigating the properties of the
phase diagram in the neighborhood of the origin.)
As may be shown by the analysis of noninstanton solutions \cite{jst},
if a potential which admits a near-to-limit instanton is less curved
than the limit potential, it necessarily admits another instanton of the
same order, with the scalar field extended along a finite segment of
the barrier. The analysis of the phase diagram as well as numerical
calculations performed for the second order instantons suggest that the
extended instanton has less action than the compact one, but as both
instantons first appear in the course of decreasing of $\xi$ they have
greater action than the HM instanton. The crucial question is which
instanton has the least action. We have shown that (1) for potentials
with $\xi \lesssim - 4$ and $\zeta < \zeta_{crit}$, the instanton
with the least action is the CdL instanton of the first order if
the potential admits no other CdL instanton, (2) for quartic
potential with $\Delta = 0.5$ and $\xi < -4$, the same is true
also if the potential admits two instantons of the second order.
Note that the order of the instanton determines
the number of bubbles and walls filled with the scalar field on the
true-vacuum side of the barrier which are created in the process of
quantum mechanical tunneling. For an instanton of the first order
the outcome of the tuneling is one bubble and no wall, while
for instantons of the second order the outcome is either no bubble
and one wall or two bubbles and no wall. Thus, in both cases cited
above the vacuum decays via formation of a single bubble.


\noindent
{\it Acknowledgement.} This work was supported by the grant VEGA
1/0250/03.



\appendix
\section{The second order contributions to $u$ and $v$}
\setcounter{equation}{0}
\renewcommand{\theequation}{A-\arabic{equation}}

To derive the expression (\ref{eq:Dxi2}) for $\Delta \xi$ we need to know the
functions $u_2$ and $v_2$.
The equation for $u_2$ solves for $l = 1$ by the Ansatz
$u_2 = C_1 + C_2 c^2$. (Our definition of $A$ requires that $u_2 (0) = 0$,
which can be fulfilled only if we include into $u_2$ also a term proportional
to $c$. However, since this term may be absorbed into $u_1$ it has no effect
on the results and may be ignored.) The solution is
\begin{equation}
u_2 = \frac {A^2}{12} \eta \left( \frac 12 - c^2 \right).
\label{eq:ph2s}
\end{equation}
The equation for $v_2$ obtained from the second order equation
for $v$ reads
\begin{equation}
{\cal B}^{(2)} v_2 = - \kappa ({P_l'}^2 - 2P_l^2)
s,\ \ \ {\cal B}^{(2)} = \frac {d^2}{d\chi^2} + 1.
\label{eq:al2}
\end{equation}
This is supplemented by the initial conditions $v_2 =
v_2' = 0$ at $\chi = 0$. For $l = 1$, the right hand side is
proportional to
$(s^2 - 2c^2)s = (1 - 3c^2)s$, which suggests the Ansatz $v_2
= (C_1 + C_2 c^2)s + C_3 \chi c$. (The term proportional to $c$ is
omitted because of the
first initial condition.) If we insert this Ansatz into the equation
for $v_2$ and use the second initial condition we obtain
\begin{equation}
v_2 = \frac 14 \left[ \left( 1 - \frac 32 c^2\right)s +
\frac 12 \chi c\right].
\label{eq:al2s}
\end{equation}
This provides the expressions for $\delta$ and $\hat \delta$
cited in section~\ref{sec:B34}. The latter expression is valid also
for higher values of $l$ since $v_2$ always splits into a part
containing $s$ and a term proportional to $\chi c$. Note that our
expressions for $u_2$ and $v_2$ are consistent with the expansions
of $a$ and $\phi$ in \cite{sta}.

The expression (\ref{eq:B40}) for $B_4$ allows us to avoid an
explicit computation of this quantity once we have computed $\Delta
\xi$. For completeness, let us explain how the part of (\ref{eq:B40})
containing $v_2$ is obtained. The starting expression is
$$\int_0^\pi \left[ Q v_2 - \frac 2\kappa (2c v_2 v_2' + s {v_2'}^2 -
3s v_2^2) \right] d\chi.$$
Rewrite the expression in the round brackets as
$$(c v_2^2)' + (s v_2 v_2')' - 2s v_2^2  - (s v_2')' v_2.$$
The integral of the first two terms over $\chi$ equals
$- \delta^2$ which leads to the expression for $b_4^{int}$
cited in section~\ref{sec:B34}, while the remaining two terms may be
rewritten as $v_2 {\cal B} v_2$. The equation for $v_2$ resulting from the
variation of $B_4$ (the second equation (\ref{eq:ph2al2})) can
be obtained also by combining the first order equation for $v_2$,
\begin{equation}
{\cal B}^{(1)} v_2 = - \frac 12 \kappa \left( \frac 12 {P_l'}^2
+ 2 P_l^2 \right) s^2,\ \ \ {\cal B}^{(1)} = c
\frac {d}{d\chi} + s,
\label{eq:al1}
\end{equation}
with the equation (\ref{eq:al2}). For that purpose, note that
${\cal B} = - s {\cal B}^{(2)} - {\cal B}^{(1)}$.




\begin{thebibliography}{99}

\bibitem{old} A. H. Guth, Phys. Rev. {\bf D23}, 347 (1981).

\bibitem{bgt} M. Bucher, A. S. Goldhaber and N. Turok, Phys. Rev. {\bf D52},
3314 (1995); K. Yamamoto, M. Sasaki and T. Tanaka, Astrophys. J. {\bf 455},
412 (1995); A. D. Linde, Phys. Lett. {\bf B351}, 99 (1995);
A. D. Linde and A. Mezhlumian, Phys. Rev. {\bf D52}, 6789 (1995).

\bibitem{cdl} S. Coleman and F. de Luccia, Phys. Rev. {\bf D21}, 3305 (1980).

\bibitem{hmo} S. W. Hawking and I. G. Moss, Phys. Lett. {\bf 110B},
35 (1982).

\bibitem{lst} A. D. Linde: {\it Particle Physics and Inflationary Cosmology},
Harwood, Chur, Switzerland (1990).

\bibitem{hhe} S. W. Hawking and T. Hertog, Phys. Rev. {\bf D66},
123509 (2002).

\bibitem{tan} T. Tanaka, Nucl. Phys. {\bf B 556}, 373 (1999).

\bibitem{2Dt} T. Banks, C. Bender and T. T. Wu, Phys. Rev. {\bf D8},
3346 (1973); 3366 (1973).

\bibitem{par} S. Parke, Phys. Lett. {\bf B121}, 313 (1983).

\bibitem{sam} D. A. Samuel and W. A. Hiscock, Phys. Rev. {\bf D44},
3052 (1991).

\bibitem{sst}  M. Sasaki, E. D. Stewart and T. Tanaka, Phys. Rev. {\bf D50},
941 (1994).

\bibitem{sta} T. Tanaka and M. Sasaki, Progr. Theor. Phys. {\bf
88}, 503 (1992).

\bibitem{jst} L. Jensen and P. J. Steinhardt, Nucl. Phys. {\bf B237},
176 (1984); V. Balek and M. Demetrian, Phys. Rev. {\bf D69},
063518 (2004).

\bibitem{vil} N. J. Vilenkin, {\it Special Functions and the Theory of Group
Representations}, Providence, RI: Amer. Math. Soc. (1968).

\bibitem{lli} L. D. Landau and E. M. Lifshitz, {\it Mechanics (Course of
Theoretical Physics)}, Butterworth-Heineman (1995).

\end{thebibliography}
\end{document}